\documentclass[11pt]{article}
\newsavebox{\PSLASH}
\sbox{\PSLASH}{$p$\hspace{-1.8mm}/}

\input{amssym}
\def\be{\begin{equation}}
\def\ee{\end{equation}}
\def\ba{\begin{eqnarray}}
\def\ea{\end{eqnarray}}
\begin{document}
\title{\large \bf Gravitational Collapse of a
Rotating Cylindrical Null Shell in the Cosmic String Spacetime}
\author{ Samad Khakshournia$^{1}$
\footnote{Email Address: skhakshour@aeoi.org.ir}\\
$^{1}$ Nuclear Science and Technology Research Institute
(NSTRI),Tehran, Iran }\maketitle
\[\]
 \[\]
 \[ \]
 \begin{abstract}
We study the gravitational collapse of a rotating  cylindrical
null shell with flat interior and the metric of a spinning cosmic
string as the exterior. We see that there is a critical radius,
where the energy density of the shell vanishes and beyond which it
would be negative, thereby signaling that the matching would be
unphysical.
\end{abstract}
 \hspace{1.0cm}\textit{keywords}: gravitational
collapse; null shell; closed timelike curves.

\section{Introduction}
Relativistic dynamics of cylindrical symmetric thin shells as
sources of gravitational field have been studied during the recent
development of general relativity. Such objects are used as
idealized models to investigate non-spherical gravitational
collapse and the non-linearity of the field equations. A number of
papers have been concerned with stationary and non-stationary
rotating cylindrical shells in general relativity (see
Refs.\cite{gersl,Pereira} and references therein). In particular,
by considering the collapse of a cylindrical shell made of
pressure-free counter-rotating particles in vacuum, the authors
showed that even a small amount of rotation can halt the collapse
at some minimal non-zero radius.\cite{Torn} Recently, Mena et
al.\cite{Mena} considered the gravitational collapse of a rotating
cylindrical shell in vacuum. Using the matching conditions between
a Minkowski interior and the spinning cosmic string exterior
through a collapsing, rotating cylindrical shell of null dust,
they found that the shell with positive energy density bounces
before closed timelike curves (CTCs) can be created. However, it
should be noted that they examined the matching conditions across
a timelike shell and
so applied the Darmois-Israel formalism in their paper.\\
In this paper we consider the configuration studied in
Ref.\cite{Mena} from a different perspective by looking at a
rotating shell whose history is a null hypersurface. For this
purpose we use Barrab\`{e}s-Israel null shell formalism\cite{Bar1}
( see also Ref.\cite{hogen} and references therein) to investigate
the matching conditions.
\par
Natural geometrized units, in which $G=c=1$ are used throughout
the paper. The null hypersurface is denoted by $\Sigma$. The
symbol $|_{\Sigma}$ means ``evaluated on the null hypersurface ".
We use square brackets [F] to denote the jump of any quantity F
across $\Sigma$. Latin indices range over the intrinsic
coordinates of $\Sigma$ denoted by $\xi^{a}$, and Greek indices
over the coordinates of the 4-manifolds.

\section{Null Shell Formalism }

Consider the gravitational collapse of a rotating  cylindrical
shell in vacuum. We imagine that during the late stages of the
collapse, the shell tends to fall at nearly the speed of light
such that the history of the shell coincides with a null
hypersurface $\Sigma$. Taking spacetime to be flat Minkowski
inside the shell ($\cal M_{-})$, we write the metric there as
\begin{equation}\label{metricint}
ds^{2}_{-}=-dt^{2}+d\rho^{2}+\rho^{2}d\varphi^{2}+dz^{2},
\end{equation}
in terms of the Einstein-Rosen canonical cylindrical coordinates
$(t,\rho,\varphi,z)$. For the exterior vacuum spacetime of the
shell ($\cal M_{+})$, we take the metric of a spinning cosmic
string in the following form\cite{Mena,Soleng}
\begin{equation}\label{metricext}
ds^{2}_{+}=-(dT+md\Phi)^{2}+C^2dr^2+r^2d\Phi^2+dz^2,
\end{equation}
expressed in the same coordinate $z$  but in terms of distinct
coordinates $T$,  $r$ and $\Phi$ in general. The parameters $C$
and $m$ are associated with the mass per unit length and angular
momentum per unit length of the string, respectively, such that
for positive mass $C>1$. An inspection of Eq. (\ref{metricext})
shows that the region for which $r<m$ contains CTCs.\\
As seen from $\cal M_{+}$, one can obtain the description of the
hypersurface by solving the Euler-Lagrange equations. Defining
\begin{equation}\label{trans2}
L\equiv\frac{1}{2}\left(-(\dot{T}+m\dot{\Phi})^2+C^2\dot{r}^{2}
+r^2\dot{\Phi}^{2}\right),
\end{equation}
where $\dot{q}$ denotes $\frac{dq}{d\lambda}$, with $\lambda$
being an affine parameter on the null generators of the
hypersurface, the Euler-Lagrange equations are written as
\begin{equation}\label{trans3}
\frac{\partial L}{\partial q}-\frac{d}{d\lambda}\frac{\partial
L}{\partial\dot{q}}=0,
\end{equation}
where $q$ represents the coordinates $x^{\mu}_{+}=(T,r,\Phi,z).$
Now, equations for the null geodesics are found as\cite{Mena}
\begin{equation}\label{trans4}
\hspace*{0.1cm}\frac{\partial
L}{\partial\dot{T}}=-E\hspace*{0.6cm}\Longrightarrow
\hspace*{0.3cm}\dot{T}+m\dot{\Phi}=E,
\end{equation}
\begin{equation}\label{trans5}
\hspace*{0.8cm}\frac{\partial
L}{\partial\dot{\Phi}}=K\hspace*{1.0cm}\Longrightarrow
\hspace*{0.7cm}r^2\dot{\Phi}=K+mE,
\end{equation}
\begin{equation}\label{trans6}
\hspace*{3.0cm}L=0\hspace*{1.5cm}\Longrightarrow\hspace*{0.6cm}
C^2r^2\dot{r}^2=E^2r^2-(K+mE)^2,
\end{equation}
where $E$ and $K$ are constants, and it is assumed that the null
geodesics are future-directed ($E>0$) and rotating in the positive
direction such that $\dot{\Phi}>0$. From (\ref{trans6}) it is seen
that the shell always has a turning point located at
\begin{equation}\label{trans66}
b=\frac{K}{E}+m.
\end{equation}
Solving the Eqs. (\ref{trans4})-(\ref{trans6}),we get
\begin{equation}\label{trans55}
\frac{dr}{dT}=-\frac{r\sqrt{r^2-b^2}}{C(r^2-mb)}
\end{equation}
Integrating Eq. (\ref{trans55}) gives the equation of $\Sigma$ in
$\cal M_{+}$ as
\begin{equation}\label{trans77}
v_{+}\equiv T+ r^{*}=0,
\end{equation}
where $r^{*}=\int\frac{C(r^2-mb)}{r\sqrt{r^2-b^2}}dr=
C\sqrt{r^2-b^2}-mC\sec^{-1}\left(\frac{r}{b}\right)$. From the
Eqs. (\ref{trans4})-(\ref{trans6}), we can obtain the
tangent-normal vector of $\Sigma$ as viewed from $\cal M_{+}$
\begin{equation}\label{trans7}
n^{\mu}_{+}\equiv\frac{dx^{\mu}_{+}}{d\lambda}=
\left(\frac{E(r^2-mb)}{r^2},\frac{-E\sqrt{r^2-b^2}}{Cr},\frac{bE}{r^2}
,0\right)\big|_{\Sigma}.
\end{equation}
We also note that the integration of
$\frac{dr}{d\lambda}=-\frac{E}{Cr}\sqrt{r^2-b^2}$ yields
\begin{eqnarray}\label{intrinmatch}
\lambda=-\frac{C}{E}\sqrt{r^2-b^2}\big|_{\Sigma}.
\end{eqnarray}
Furthermore, from Eqs. (\ref{trans5}) and (\ref{trans6}) one gets
\begin{equation}\label{trans9}
\frac{d\Phi}{dr}=\frac{\dot{\Phi}}{\dot{r}}=
\frac{-bC}{r\sqrt{r^2-b^2}}\big|_{\Sigma}.
\end{equation}
Integration  of  Eq. (\ref{trans9}) leads to
\begin{eqnarray}\label{trans77}
\psi=\Phi+C\sec^{-1}\left(\frac{r}{b}\right)\big|_{\Sigma},
\end{eqnarray}
where $\psi$ is a constant on the null generators. We thus take
$\xi ^{a}=(\lambda,\psi,z)$ as the intrinsic coordinates on
$\Sigma$, and as these are well adapted to the generators we can
form the tangent basis vectors $e^{\mu}_{\lambda}=n^{\mu}$,
$e^{\mu}_{\psi}=\delta^{\mu}_{\phi}$, and
$e^{\mu}_{z}=\delta^{\mu}_{z}$. Now, by virtue of
$n_{\mu}e^{\mu}_{\psi}\big|_{+}=0$, we get $K=0$, so that Eq.
(\ref{trans66}) reduces to the form $b=m$. The shell 's intrinsic
metric induced from $\cal M_{+}$ is found as
\begin{equation}\label{junctioncondp}
ds^2_{+}\big|_{\Sigma}=\frac{\lambda^2E^2}{C^2}d\psi^2+dz^2.
\end{equation}
The null hypersurface $\Sigma$ as seen from $\cal M_{-}$ is
described by the parametric equations
\begin{eqnarray}\label{par1}
t+\rho&\equiv&\upsilon_{-}=const,\nonumber\\
\rho&=&-\lambda,\nonumber\\
\varphi &=& \psi, \\
z&=&z. \nonumber
\end{eqnarray}
The tangent-normal vector to $\Sigma$ as viewed from $\cal M_{-}$
is
\begin{equation}\label{junctioncondp1}
n^{\mu}_{-}\equiv\frac{dx^{\mu}_{-}}{d\lambda}=
\left(1,-1,0,0\right)\big|_{\Sigma}.
\end{equation}
It is then seen that the induced metric on $\Sigma$ from $\cal
M_{-}$ is given by
\begin{equation}\label{junctioncondm}
ds^2_{-}\big|_{\Sigma}=\lambda^2d\psi^2+dz^2.
\end{equation}
Now from (\ref{junctioncondp}) and (\ref{junctioncondm}), the
requirement of continuity of the induced metric on $\Sigma$ yields
the following matching condition
\begin{equation}\label{junction1}
E=C.
\end{equation}
We may then complete the basis by a transverse null vector
$N^{\mu}$ uniquely defined by the four conditions
$n_{\mu}N^{\mu}=-1$, $N_{\mu}e^{\mu}_{A}=0$ $(A=\psi,z)$, and
$N_{\mu}N^{\mu}=0$. We find for both sides
\begin{eqnarray}\label{normaltrans}
N_{\mu}|_{-}&=&\frac{1}{2}\left(-1,+1,0,0\right)\big|_{\Sigma},\\
N_{\mu}|_{+}&=&\frac{1}{2C}\left(\frac{-r^2}{r^2-m^2},\frac{Cr}
{\sqrt{r^2-m^2}},0,0\right)\big|_{\Sigma}.
\end{eqnarray}
To proceed further, we here need to define  a pseudo-inverse of
the induced metric $g_{ab}$ on $\Sigma$ as $g_{*}^{ac}g_{bc} =
\delta_{b}^{a}+ n^{a}N_{\mu}e^{\mu}_{b}$, with
$n^{a}=\delta^{a}_{\lambda}$,\cite{Bar1} leading to
$g^{ab}_{*}$=diag$\left(0,\frac{1}{\lambda^{2}},1\right)$.\\ The
final junction condition is formulated in terms of the jump in the
extrinsic curvature. Using the definition ${\cal
K}_{ab}=e^{\mu}_{a}e^{\nu}_{b}\nabla_{\mu}N_{\nu}$, we can
therefore compute the transverse extrinsic curvature
tensor\cite{Bar1} on both sides of $\Sigma$. Its non-vanishing
component on the minus side is found as
\begin{equation}\label{K33m}
{\cal K}_{\psi\psi}|_{-}=\frac{\rho}{2}\big|_{\Sigma}.
\end{equation}
The corresponding non-vanishing components on the plus side are
computed as
\begin{equation}\label{Ktetteta2}
{\cal
K}_{\psi\psi}|_{+}=\frac{r^2}{2C^2\sqrt{r^2-m^2}}\big|_{\Sigma},
\end{equation}
\begin{equation}\label{K11p}
{\cal K}_{\lambda\psi}|_{+}=\frac{m}{C\sqrt{r^2-m^2}}.
\end{equation}
The surface energy-momentum tensor of the lightlike shell having
the null hypersurface $\Sigma$ as its history is directly related
to the jump in the transverse extrinsic curvature. In the tangent
basis $e_{a}$, it can be written in the form\cite{Pois}
\begin{equation}\label{juncnullintr}
S^{ab}=\mu n^{a}n^{b}+pg_{*}^{ab}+j^{a}n^{b}+j^{b}n^{a},
\end{equation}
where
\begin{equation}\label{nullenergy}
\mu=-\frac{1}{8\pi}g_{*}^{ab}[{\cal K}_{ab}]
\end{equation}
represents the surface energy density,
\begin{equation}\label{nullpressure}
p=-\frac{1}{8\pi}[{\cal K}_{ab}]n^{a}n^{b}
\end{equation}
displays the isotropic surface pressure, and
\begin{equation}\label{nullcurrent}
j^{a}=-\frac{1}{8\pi}g_{*}^{ac}[{\cal K}_{cd}]n^{d}
\end{equation}
represents the surface current of the lightlike shell. All these
surface quantities are measured by a family of freely-moving
observers crossing the null hypersurface. Using the jumps in the
extrinsic curvature obtained above, we first notice that the
surface pressure term vanishes identically. The energy density and
surface current are then calculated as
\begin{eqnarray}\label{densnull}
\mu=\frac{(C^2-1)r^2-C^2m^2}{16\pi
C^2(r^2-m^2)^{\frac{3}{2}}}\big|_{\Sigma},
\end{eqnarray}
\begin{eqnarray}
\label{pressrnull} j^{\psi}=\frac{m}{8\pi
C(r^2-m^2)^{\frac{3}{2}}}\big|_{\Sigma}.
\end{eqnarray}
From (\ref{densnull}), it is seen that due to the interplay of
both gravitational and centrifugal energies, which enter in the
expression (\ref{densnull}) for the energy density of the shell
with opposite signs and different dependence on the shell radius,
there is a critical radius where the surface energy density of the
shell becomes zero. It is given by
\begin{equation}\label{nullpressure1}
r_{c}=\frac{mC}{\sqrt{C^2-1}}.
\end{equation}
For $r<r_{c}$, the energy density $\mu$ is always negative, hence,
in order to avoid unphysical negative energy densities, as
Dray\cite{Dray} has suggested, it is more natural to expect that
the shell starts expanding at the moment when $\mu$ becomes zero,
but the physical mechanism by which the shell bounces at this
critical radius, is unknown. Note that with $C>1$, from
(\ref{nullpressure1}) it is clear that $r_{c}>m$, meaning that the
energy density of the shell starts becoming negative  before the
radius at which CTCs would be created, can be reached.

\section{Conclusion}

In this paper, we have used Barrab\`{e}s-Israel null shell
formalism to study the relativistic dynamics of a collapsing
rotating cylindrical null shell with flat interior and a spinning
cosmic string exterior spacetimes. We have seen that the energy
density of the shell inevitably becomes negative at radii less
than a critical radius given by (\ref{nullpressure1}). It turns
out that this radius is larger than the radius at which CTCs can
be formed in the exterior. Hence, this idealized model, as a test
bed for the cosmic censorship conjecture, due to the lack of a
physically reasonable distributional matter content on the shell
does not rule out this idea. It is remarkable that how our results
differ from those of reported in Ref.\cite{Mena} for a collapsing
shell of null dust whose history is a timelike hypersurface and
according to the Darmois-Israel matching conditions this shell
with the positive definite energy density bounces before CTCs
arise.

\end{document}